\pdfoutput=1
\documentclass[aps,twocolumn,floatfix,prb,showpacs]{revtex4}
\usepackage{graphicx,amsmath,bbm}

\newcommand{\ket}[1]{{| #1 \rangle}}
\newcommand{\bra}[1]{{\langle #1 |}}

\newcommand{\ua}{\uparrow}
\newcommand{\da}{\downarrow}

\def\braket#1{\mathinner{\langle{#1}\rangle}}
\def\bra#1{\left\langle#1\right|}
\def\ket#1{\left|#1\right\rangle}

\begin{document}

\title{Spin-flip phonon-mediated charge relaxation in double quantum dots}
\date{\today}

\author{J.\ Danon}
\affiliation{Dahlem Center for Complex Quantum Systems and Fachbereich Physik, Freie Universit\"{a}t Berlin, Arnimallee 14, 14195 Berlin, Germany}
\affiliation{Niels Bohr International Academy, Niels Bohr Institute, University of Copenhagen, Blegdamsvej 17, 2100 Copenhagen, Denmark}

\begin{abstract}
We theoretically study the $(1,1)$ triplet to $(0,2)$ singlet relaxation rate in a lateral gate-defined double quantum dot tuned to the regime of Pauli spin blockade. We present a detailed derivation of the effective phonon density of states for this specific charge transition, keeping track of the contribution from piezoelectric as well as deformation potential electron-phonon coupling. We further investigate two different spin-mixing mechanisms which can couple the triplet and singlet states: a magnetic field gradient over the double dot (relevant at low external magnetic field) and spin-orbit interaction (relevant at high field), and we also indicate how the two processes could interfere at intermediate magnetic field. Finally, we show how to combine all results and evaluate the relaxation rate for realistic system parameters.
\end{abstract}


\maketitle

\section{Introduction}

The last decade has seen a great interest in spin qubits hosted in semiconductor quantum dots, motivated by the prospects of easy scalability, weak coupling to external perturbations, and flexible tunability.\cite{qdqb,ronaldrev} Experimental advance has been substantial in the past years, and essential operations including qubit initialization, manipulation, and readout have been convincingly demonstrated.\cite{frank:nature,laird,pioro:nature} This progress is not only exciting in the context of quantum computation and information, but it also provides a unique platform for studying fundamental quantum properties of nanoscale systems. Ongoing effort is therefore directed at improving the quality of the spin qubits, mainly by trying to reduce qubit dephasing\cite{bluhm:natphys} and increase the measurement fidelity.\cite{PhysRevB.85.035306}

A common method to read out a quantum dot spin qubit relies on the so-called Pauli spin blockade:\cite{ono:science} A double quantum dot is tuned to a $(1,1)$ charge state, meaning that each dot contains exactly one excess electron,\cite{endnote1} and two of the four resulting $(1,1)$ spin states are used as a qubit basis. After qubit manipulation, the double dot potential is tilted such that a $(0,2)$ charge state becomes the two-electron ground state. For not too strong tilting, the only accessible $(0,2)$ state is a spin singlet, which makes the $(1,1)\to (0,2)$ charge relaxation spin-selective. If the two qubit basis states contain a different spin-singlet component, then one can use charge detection to measure the qubit's final state before tilting, either by doing transport measurements coupling the doubly occupied dot strongly to an outgoing lead,\cite{frank:science,frank:nature} or by detecting the charge state with a nearby charge sensor.\cite{johnson:nature,petta:067601,reillyt2}

The accuracy of such a readout depends crucially on the effectiveness of the spin blockade: Any leakage out of the blocked triplet states reduces the readout visibility and thereby distorts the measurement.\cite{PhysRevB.85.035306} A detailed understanding of the spin-flip relaxation responsible for such triplet leakage is thus essential in the context of spin qubit measurement. Most existing theoretical work along these lines was done for {\it single-dot} spin relaxation\cite{khaetskii2,PhysRevLett.98.126601,PhysRevB.77.115438} or consists of numerical studies of the relaxation rates.\cite{stano:arxiv,PhysRevB.86.205321} A thorough analytical study of interdot spin-flip charge relaxation in double quantum dots is still missing.

Here, we study in detail the $(1,1)$ triplet to $(0,2)$ singlet decay rate for a lateral double quantum dot.
We investigate two spin-mixing mechanisms which provide a coupling between the otherwise orthogonal states: (i) For small externally applied magnetic fields the coupling is believed to be dominated by the effective magnetic field gradient over the two dots caused by the hyperfine coupling of the electron spins to the randomly fluctuating nuclear spins in the host material.\cite{johnson:nature,PhysRevB.85.035306} (ii) At larger fields, this coupling is suppressed for the two polarized triplet states for which spin-orbit interaction takes over as dominating spin-mixing mechanism.\cite{larsnatcomm}
The second ingredient necessary for a finite leakage rate is the dissipation of the energy difference $\Delta$ between the initial $(1,1)$ triplet and final $(0,2)$ singlet state, which we assume to be provided by the coupling to acoustical phonons in the host material. We derive the function $P_1(\Delta)$ for this specific charge transition, which gives the probability that the transition is accompanied by the dissipation of energy $\Delta$ by a single phonon (alternatively one could call this function the effective phonon density of states for the charge transition). We include the contribution from piezoelectric as well as deformation potential electron-phonon coupling, and we find that the piezoelectric contribution to $P_1(\Delta)$ is linear at low energies and $\propto \Delta^{-5}$ at high energies, whereas the coupling to the deformation potential leads to a contribution $\propto \Delta^3$ for low energies and $\propto \Delta^{-1}$ for high energies.
We finally evaluate explicit relaxation rates using parameters of the experiment of Ref.\ \onlinecite{larsnatcomm}. We find a decay rate $\sim$~MHz, which agrees with experimental observations.\cite{florist1}

Our results are not only relevant for spin qubit readout. In Ref.\ \onlinecite{larsnatcomm} it was suggested that the spin-orbit coupling of the $(1,1)$ triplet and $(0,2)$ singlet states could also be utilized to drive off-resonant microwave-stimulated Raman transitions within the $(1,1)$ space. In that case transitions between the $(1,1)$ and $(0,2)$ states would contribute to qubit dephasing, and a detailed understanding of the mechanisms responsible for these transitions would be essential in this context as well.

The rest of this paper is structured as follows. In Sec.~\ref{sec:model} we introduce our model of the double quantum dot and present an effective Hamiltonian defining the basis we will work in. In Sec.\ \ref{sec:smpert} we then investigate the two spin-mixing mechanisms (spin-orbit interaction and a magnetic field gradient over the double dot) and we derive the matrix elements needed to calculate the leakage rate. In Sec.\ \ref{sec:eph} we study the coupling to the phonon bath in detail. We start from the standard Hamiltonian describing the electron-phonon coupling and derive from it $P_1(\Delta)$ for the $(1,1)$ to $(0,2)$ spin-flip charge transition. Finally, in Sec.~\ref{sec:rel}, we evaluate the leakage rate explicitly with realistic experimental parameters.

\section{Model}\label{sec:model}

\begin{figure}[b]
 \includegraphics{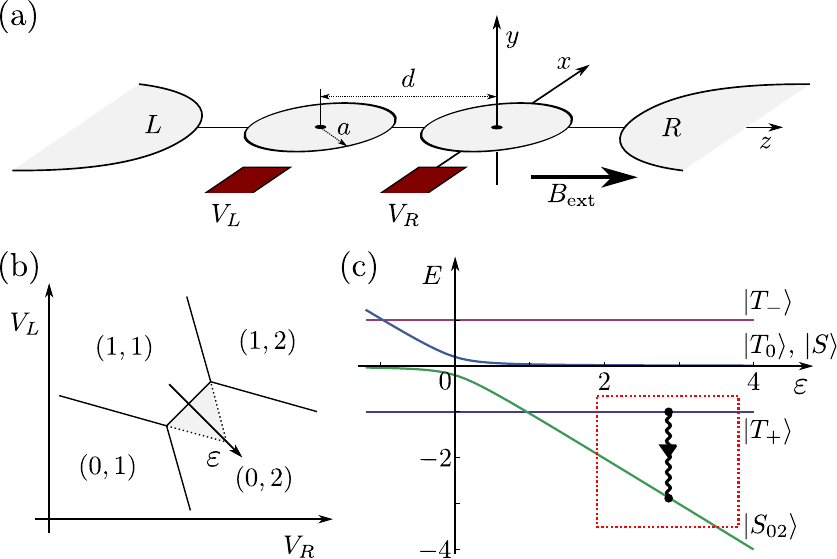}
 \caption{(a) A lateral double quantum dot tunnel coupled to a left and right lead. Two nearby gate electrodes with applied voltages $V_{L,R}$ can change the potential offset of the dots. An in-plane external magnetic field $B_{\rm ext}$ is applied in a direction parallel to the double-dot axis. (b) Charge stability diagram of a few-electron double dot. We investigate the gray region, where both $(1,1)$ and $(0,2)$ can be stable charge states. (c) The spectrum of the electronic Hamiltonian (\ref{eq:ham0}) as a function of the detuning $\varepsilon$, where we chose $t=0.2$ (all energies in units of $B_{\rm ext}$). The detuning axis $\varepsilon$ is also indicated in (b). We will focus on the regime indicated with the red rectangle, where the two lowest-energy states are $\ket{S_{02}}$ and $\ket{T_+}$.}
 \label{fig:dqd}
\end{figure}
We consider a lateral double quantum dot, gate-defined in a two-dimensional electron gas (2DEG) formed at the interface of an AlGaAs/GaAs heterostructure which has been grown in the crystallographic (001) direction. The setup we have in mind is sketched in Fig.\ \ref{fig:dqd}a: The two dots (gray circular areas) are separated by a distance $d$. Each dot is approximated to be a two-dimensional harmonic potential well with a typical width $a$, corresponding to a single-dot level spacing $\hbar\omega = \hbar^2/2ma^2$ where $m$ is the effective electron mass ($m \approx 0.067m_{\rm e}$ in GaAs). Nearby gate electrodes with applied voltages $V_L$ and $V_R$ couple capacitively to the dots and can change the potential offset of the two dots. We choose the $z$-axis to be parallel to the interdot axis and assume an external magnetic field $B_{\rm ext}$ to be applied in the same $z$-direction (as was the case in the experiment of Ref.\ \onlinecite{larsnatcomm}). Two nearby leads, labeled $L$ and $R$, are tunnel coupled to the two dots and act thereby as electron reservoirs. 

Part of the few-electron charge stability diagram of such a double dot is shown in Fig.\ \ref{fig:dqd}b. In the plane $(V_L,V_R)$ regions of stable charge configuration $(n_L,n_R)$ have a hexagonal shape, where $n_{L(R)}$ denotes the number of excess electrons on the left(right) dot.\cite{vanderwielrev} The regime we want to investigate in this work is the shaded triangle close to the $(1,1)$-$(0,2)$ boundary. Here the ground state of the system is a $(0,2)$ charge state, but excited $(1,1)$ states are also stable in the sense that they cannot decay through sequential tunneling processes like $(1,1) \to (0,1) \to (0,2)$. We also assume the tunnel barriers to the leads to be high enough that the corresponding cotunneling processes do not contribute significantly to charge relaxation. In that case, we can regard the double dot to be a closed system containing two electrons which have either a $(1,1)$ or $(0,2)$ charge configuration.

The single-dot orbital level spacing $\hbar\omega \sim {\rm meV} \sim 10$~K is typically comparable to the charging energy of the dot, i.e., the Coulomb energy it costs to add an extra electron to the dot. This is such a large energy scale that we can disregard all electronic states involving higher orbital states and focus on electrons in the orbital ground state. Explicitly, the two-dimensional ground state wave functions in the left and right potential wells read $\psi_L({\bf r}) = (1/\sqrt{2\pi} a)\exp \{-[x^2+(z+d)^2]/4a^2\}$ and $\psi_R({\bf r}) = (1/\sqrt{2\pi} a) \exp\{-[x^2+z^2]/4a^2\}$. Using a Hund-Mulliken approach we orthonormalize these two wave functions, resulting in the basis states\cite{burkard_prb}
\begin{align}
\Psi_{L(R)}({\bf r}) = \frac{\psi_{L(R)}({\bf r}) - g\psi_{R(L)}({\bf r})}{\sqrt{1-2gs+g^2}},
\end{align}
with the factor $g = (1-\sqrt{1-s^2})/s$ and the overlap integral $s = \int d{\bf r}\,\psi^*_L({\bf r})\psi_R({\bf r}) = \exp \{-d^2/8a^2 \}$.

Including spin into the picture, we can construct from these basis states a $(1,1)$ spin-singlet state $\ket{S}$, three $(1,1)$ spin-triplet states $\ket{T_0}$ and $\ket{T_\pm}$, and a $(0,2)$ spin-singlet state $\ket{S_{02}}$. We assume that in the basis spanned by these five states the Hamiltonian describing the kinetic and potential energy of the two electrons as well as their Coulomb interaction can be written as\cite{burkard_prb}
\begin{align}
 \hat H = {} & \hat H_0 + \hat H_{\rm t}, \label{eq:ham0}\\
 \hat H_0 = {} & B_{\rm ext} \{ \ket{T_-}\bra{T_-} - \ket{T_+}\bra{T_+} \} - \varepsilon \ket{S_{02}}\bra{S_{02}}, \\
 \hat H_{\rm t} = {} & t \{ \ket{S}\bra{S_{02}} + \ket{S_{02}}\bra{S} \},
\end{align}
where we included a Zeeman term coupling to the spin of the two electrons. The field $B_{\rm ext}$ is written in units of energy (the sign chosen reflects the fact that the $g$-factor in GaAs is negative, so that a positive $B_{\rm ext}$ corresponds to a positive field along the $z$-axis) and $\varepsilon$ describes the detuning between $\ket{S_{02}}$ and the unpolarized $(1,1)$ states.\cite{endnote2} Increasing $\varepsilon$ can be effected by changing $V_L$ and $V_R$ as indicated with the arrow in Fig.\ \ref{fig:dqd}b.

In Fig.\ \ref{fig:dqd}c we plot the spectrum of $\hat H$ as a function of the detuning $\varepsilon$, where we chose $t = 0.2$ (all energies in units of $B_{\rm ext}$). Our regime of interest (the gray area in Fig.\ \ref{fig:dqd}b) is where $\ket{S_{02}}$ is the ground state, indicated with the red dotted rectangle in Fig.\ \ref{fig:dqd}c. We assume that in this regime $\varepsilon \gg t$, so that the tunneling Hamiltonian $\hat H_{\rm t}$ can be treated as a perturbation. An excited $(1,1)$ state can only decay to the $(0,2)$ ground state if its wave function is a spin-singlet (or contains a singlet component). If the system is in a pure $(1,1)$ spin-triplet, it cannot decay and stays blocked in the excited state.

We now focus on such a spin-blocked situation and assume that the system is initially in $\ket{T_+}$. The purpose is to calculate the relaxation rate from $\ket{T_+}$ to $\ket{S_{02}}$ (see the wiggly line in Fig.\ \ref{fig:dqd}c). For this we need two ingredients, which we will investigate in detail in the next two Sections: (i) We need a perturbation $\hat H_{\rm sm}$ with a finite matrix element between the states $\ket{T_+}$ and $\ket{S_{02}}$. (ii) The energy difference between initial and final state $E_{T_+} - E_{S_{02}}\equiv \Delta$ has to be dissipated by the environment of the double dot, for which we assume the coupling to acoustic phonons to be responsible.

\section{Spin-mixing perturbations}\label{sec:smpert}

As noted before, $\ket{T_+}$ and $\ket{S_{02}}$ are orthogonal in spin space, and the perturbation $\hat H_{\rm sm}$ thus has to be of a spin-mixing nature. We will investigate two such perturbations: (i) spin-orbit interaction, which mixes the spin and orbital parts of the electrons' wave functions, and (ii) a magnetic field gradient over the double dot, i.e., a difference between the effective magnetic fields at the positions of the left and right dot. Below we will introduce the two perturbations and calculate the resulting matrix elements coupling $\ket{T_+}$ and $\ket{S_{02}}$.

\subsection{Spin-orbit interaction}

Spin-orbit interaction perturbs single-particle states in the two dots resulting in mixed spin-orbital eigenstates instead of pure spin states.\cite{ronaldrev} Thereby it can give rise to ``spin-flip'' tunnel coupling of states with apparent opposite spin.\cite{PhysRevB.80.041301,PhysRevB.81.201305,larsnatcomm} For each electron the spin-orbit Hamiltonian reads
\begin{equation}
\hat {\tilde H}_\text{so} =  \alpha (-\hat p_{\tilde y} \hat \sigma_{\tilde x} + \hat p_{\tilde x} \hat \sigma_{\tilde y}) + \beta (-\hat p_{\tilde x}\hat \sigma_{\tilde x} + \hat p_{\tilde y} \hat \sigma_{\tilde y} ),
\label{eq:hso1}
\end{equation}
where $\hat{\bf p}$ is the momentum of the electron and $\hat \sigma_{x,y,z}$ are the three Pauli matrices. The first term in (\ref{eq:hso1}) is the so-called Rashba term and the second the Dresselhaus term. For typical 2DEG's in GaAs the corresponding spin-orbit length $l_{\rm so}$, i.e.\ the distance an electron has to travel to have its spin rotated by $\sim 1$, is of the order $l_{\rm so} \sim$ 10 $\mu$m, usually much larger than the size of the dots.\cite{ronaldrev,katja:science} The ratio of the two parameters, $\alpha/\beta$, depends on the detailed confining potential of the 2DEG and can in practice be smaller as well as larger than 1.

\begin{figure}[t]
 \includegraphics{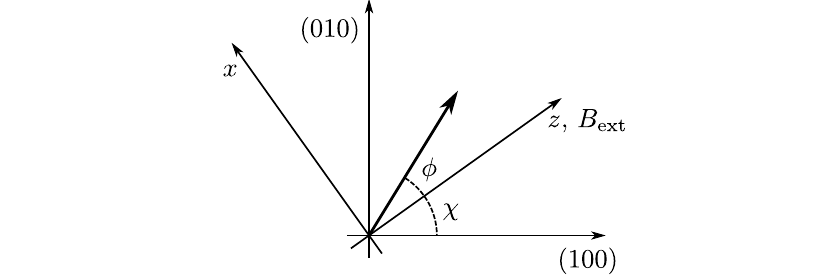}
 \caption{The orientation of our $xz$-plane with respect to the (100) and (010) crystallographic directions.}
 \label{fig:dir}
\end{figure}

The spin-orbit Hamiltonian (\ref{eq:hso1}) is written such that the $\tilde x$-, $\tilde y$-, and $\tilde z$-axes point respectively along the (100), (010), and (001) crystallographic axes, and it is assumed that the 2DEG lies in the $\tilde x\tilde y$-plane. Transforming this Hamiltonian to the coordinate system of Fig.\ \ref{fig:dqd} we find
\begin{align}
\hat H_\text{so} = {} & \alpha [-\hat p_{x} \hat \sigma_{z} + \hat p_{z} \hat \sigma_{x}]  \nonumber\\
& + \beta [\hat p_{z}(\hat \sigma_{x}\sin2\chi - \hat\sigma_z\cos2\chi) \nonumber\\
& \hspace{4em}+ \hat p_{x}(\hat \sigma_{x}\cos2\chi + \hat\sigma_z\sin2\chi)],
\label{eq:hso}
\end{align}
where $\chi$ is the angle between the double-dot axis and the (100) crystallographic direction, see Fig.\ \ref{fig:dir}.

In the two-electron position representation, the basis states $\ket{T_+}$ and $\ket{S_{02}}$ read explicitly
\begin{align*}
\braket{{\bf r}_{1,2}|T_+} & = \tfrac{1}{\sqrt{2}}\big\{ \Psi_{L\ua}({\bf r}_1)\Psi_{R\ua}({\bf r}_2)-\Psi_{R\ua}({\bf r}_1)\Psi_{L\ua}({\bf r}_2) \big\},\\
\braket{{\bf r}_{1,2}|S_{02}} & = \tfrac{1}{\sqrt{2}}\big\{ \Psi_{R\ua}({\bf r}_1)\Psi_{R\da}({\bf r}_2)-\Psi_{R\da}({\bf r}_1)\Psi_{R\ua}({\bf r}_2) \big\},
\end{align*}
where the $\Psi({\bf r})$ are now in fact two-component spinors. Using that $\hat{\bf p} = -i\hbar \partial_{\bf r}$ we find straightforwardly that only terms with $\hat p_z$ have a non-vanishing matrix element between $\ket{T_+}$ and $\ket{S_{02}}$, the total matrix element reading
\begin{equation}
 T_{\rm so}= \braket{S_{02}|\hat H_{\rm so}|T_+} = -i\hbar \frac{d}{4a^2} \frac{s}{\sqrt{1-s^2}}( \alpha + \beta\sin2\chi ).
 \label{eq:mso}
\end{equation}
We see that $T_{\rm so}$ depends non-trivially on the angle $\chi$. For a double dot with its interdot axis pointing in the crystal $(1 10)$ direction the two spin-orbit terms add constructively and $T_{\rm so}\propto(\alpha+\beta)$, whereas dots with the interdot axis along the $(\bar 110)$ direction have $T_{\rm so} \propto (\alpha-\beta)$.

We also note here that the direction of the external field $B_{\rm ext}$ can have a great influence on the effectiveness of the coupling. For instance, for a double dot with the interdot axis constructed along the $(110)$ direction (i.e., $\chi = \pi/4$), the Hamiltonian (\ref{eq:hso}) reads
\begin{align}
\hat H_\text{so} = \alpha (-\hat p_{x} \hat \sigma_{z} + \hat p_{z} \hat \sigma_{x}) + \beta (\hat p_{z}\hat \sigma_{x}+ \hat p_{x} \hat\sigma_z).
\end{align}
If the external field now points in the $x$-direction, then all terms with $\hat p_z$ come with $\hat \sigma_x$, which cannot provide a spin-flip. In this case we have thus $T_{\rm so} = 0$.

There is however another, second-order, spin-orbit mediated process which can take place: $\hat H_{\rm so}$ couples the ground state $\ket{T_+}$ to a $(1,1)$ spin-singlet which involves an excited orbital state. This excited state can be assumed to be coupled to $\ket{S_{02}}$ with a coupling energy of $\sim t$. The spin-flip now takes place inside one of the dots, and does not depend on the orientation of the interdot axis. An estimate of the magnitude of the resulting effective matrix element coupling $\ket{T_+}$ to $\ket{S_{02}}$ gives
\begin{equation}
 T_{{\rm so},2} \sim (\alpha+\beta) \frac{\hbar}{a} \frac{t}{\hbar\omega},
\end{equation}
If we assume that $t \sim \mu$eV and $\hbar\omega\sim$~meV, then a comparison with the $T_{\rm so}$ found above yields that the ``direct'' coupling dominates as long as $d/a \lesssim 8$ is the case.\cite{sorel.note} Of course, if $T_{\rm so} = 0$ (due to a special orientation of the interdot axis and $B_{\rm ext}$) then $T_{{\rm so},2}$ still provides a spin-orbit assisted spin-mixing coupling between $\ket{T_+}$ and $\ket{S_{02}}$.

\subsection{Magnetic field gradient}

A magnetic field gradient over the two dots generally mixes all four (1,1) states. All triplet states acquire a spin singlet component, and thus are coupled to $\ket{S_{02}}$ by the tunneling Hamiltonian $\hat H_{\rm t}$. Such gradients could be due to a deliberately fabricated on-chip micromagnet\cite{larsnatcomm,pioro:nature} or to the effective magnetic fields set up by the randomly fluctuating nuclear spins of the host material.\cite{frank:science,jouravlev:prl}

The Hamiltonian describing the coupling of two different effective magnetic fields $\delta{\bf B}_L$ and $\delta{\bf B}_R$ to the two electrons reads
\begin{align}
\hat H_{\rm gr} = {} & -\delta{\bf B}_L\cdot\hat{\bf S}_L -\delta{\bf B}_R \cdot\hat{\bf S}_R,
\end{align}
where we again have set $g\mu_B = -1$ for convenience, and $\hat{\bf S}_{L(R)}$ is the (dimensionless) spin operator for the electron in the left(right) dot. In the basis of the $(1,1)$ singlet and triplet states this Hamiltonian reads\cite{jouravlev:prl}
\begin{align}
\hat H_{\rm gr} = {}& \sum_\pm \left\{ -\frac{\delta B_s^\pm}{\sqrt{2}} \ket{T_0}\bra{T_\pm} \pm \frac{\delta B_a^\pm}{\sqrt{2}} \ket{S}\bra{T_\pm} + {\rm H.c.} \right\} \nonumber\\
& - \delta B_s^z \big\{ \ket{T_+}\bra{T_+}-\ket{T_-}\bra{T_-} \big\} \nonumber\\
& - \delta B_a^z \big\{\ket{S}\bra{T_0} + \ket{T_0}\bra{S}\big\},
\end{align}
where we used the notation $\delta B_{s,a}^\pm = \delta B_{s,a}^x \pm i\delta B_{s,a}^y$. The symmetric and antisymmetric fields we used are defined as $\delta {\bf B}_s = \tfrac{1}{2}(\delta {\bf B}_L + \delta {\bf B}_R)$ and $\delta {\bf B}_a = \tfrac{1}{2}(\delta {\bf B}_L - \delta {\bf B}_R)$.

We assume that the fields $\delta B_L$ and $\delta B_R$ are much smaller than the externally applied field. Then we can use first-order perturbation theory to find the singlet admixture in $\ket{T_+}$ caused by $\hat H_{\rm gr}$, which yields $\ket{T_+} \approx \ket{T_+} - (\delta B_a^+/\sqrt{2}B_{\rm ext}) \ket{S}$. Therefore, the spin-flip matrix element due to the perturbation $\hat H_{\rm t} + \hat H_{\rm gr}$ reads
\begin{equation}
 T_{\rm gr} = \braket{S_{02}|\hat H_{\rm t} + \hat H_{\rm gr}|T_+} = -t\frac{\delta B_a^+}{\sqrt{2}B_{\rm ext}}.
\label{eq:mgr}
\end{equation}

If the gradients are caused by randomly fluctuating nuclear spins, the typical magnitude of the effective fields is approximately $\delta B_{L,R} \sim A/\sqrt{N}$, where $A$ is the material-specific hyperfine coupling energy and $N$ is the number of nuclei in each dot. For GaAs dots $A \sim 100~\mu$eV and typically $N \sim 10^6$, implying that $\delta B_{L,R}$ is in the regime of 1--5 mT, which has been confirmed experimentally.\cite{frank:science}

\section{Electron-phonon coupling: Single-phonon probability}\label{sec:eph}

Both perturbations outlined above provide effectively a coupling between $\ket{T_+}$ and $\ket{S_{02}}$ and can thus cause relaxation. The only ingredient still missing to fully describe transitions between the two levels, is a mechanism dissipating the energy difference $\Delta$ between initial and final state, typically 50--500 $\mu$eV. We assume that this energy is absorbed by the (acoustical) phonon bath in the host material. Often the contribution to the electron-phonon coupling from the deformation potential is neglected, which is generally justified for phonons with energies below $\sim 10$~K.\cite{PhysRevB.71.205322} However, phonons with a wave vector larger than the inverse in-plane dot size $1/a$ are emitted almost exclusively in the $y$-direction (perpendicular to the plane of the 2DEG), and emission of piezoelectric phonons in this direction is strongly suppressed by the crystalline anisotropy.\cite{khaetskii2,PhysRevLett.98.126601} Since for a typical GaAs-hosted double quantum dot system a phonon wave vector of $1/a$ corresponds to an energy of $\sim 100~\mu$eV, we will keep in our calculation both the coupling to piezoelectric phonons and deformation phonons.

The Hamiltonian describing the coupling between the electrons (density operator $\hat\rho$) and phonons (creation and annihilation operators $\hat a^{(\dagger)}$) reads
\begin{align}
\hat H_\text{e-ph} & = \sum_{{\bf q},p} \lambda_{{\bf q},p} \hat\rho_{\bf q} [\hat a_{{\bf q},p} + \hat a_{-{\bf q},p}^\dagger] , \label{eq:hep}
\end{align}
where $\lambda_{{\bf q},p}$ are the coupling matrix elements and $\hat \rho_{\bf q} = \int d{\bf r}\, e^{-i{\bf q}\cdot{\bf r}} \hat\rho ({\bf r}) $ is the Fourier transform of the electronic density operator. The sum runs over all allowed phonon wave vectors ${\bf q}$ and includes three polarizations (one longitudinal and two transversal), labeled by $p = l,t_1,t_2$. We will neglect the mismatch of phonon velocities at the GaAs-AlGaAs interface, and treat the phonon bath as that of bulk GaAs.

The matrix elements for electron-phonon coupling read
\begin{equation}
\lambda_{{\bf q},p} = M^{(p)}_{\rm ph} \sqrt{ \frac{\hbar}{2\rho {\cal V} \omega_{{\bf q},p}}},
\end{equation}
where $\rho$ is the mass density ($\rho = 5.3 \times 10^3$ kg/m$^3$ for GaAs), ${\cal V}$ the normalization volume, and we will assume that the phonons have an isotropic linear dispersion relation at all energies of interest, i.e., $\omega_{{\bf q},p} = v_pq$, with $v_p$ the polarization-dependent sound velocity.

The constant $M^{(p)}_{\rm ph} = M^{(p)}_{\rm pe}+M_{\rm def}$ contains a contribution from both types of electron-phonon coupling,\cite{PhysRevB.48.11144}
\begin{align}
 M^{(p)}_{\rm pe} = {} & ieh_{14}A_{{\bf q},p}, \\
 M_{\rm def} = {} & \Xi q \,\delta_{p,l}, \label{eq:mdef}
\end{align}
where the coupling to the deformation potential only involves longitudinal phonons, expressed by the $\delta$-function in (\ref{eq:mdef}). We used here the piezo-electric constant, $h_{14} = 1.38\times 10^9$ V/m in GaAs, and the deformation potential, which is $\Xi = 13.7$~eV for GaAs.\cite{PhysRevB.48.11144} The coupling to piezoelectric phonons involves the anisotropy factors
\begin{equation}
A_{{\bf q},p} = \frac{2}{q^2} \left[q_{\tilde x}q_{\tilde y}e_{\tilde z}^{(p)}+q_{\tilde z}q_{\tilde x}e_{\tilde y}^{(p)}+q_{\tilde y}q_{\tilde z}e_{\tilde x}^{(p)}\right],\label{eq:anis}
\end{equation}
where ${\bf e}^{(p)}$ is the unit polarization vector for the polarization $p$. The factors as written in (\ref{eq:anis}) are in the coordinate system of the crystal structure, i.e.\ the $\tilde x$-direction along (100), $\tilde y$ along (010), and $\tilde z$ along (001). We would like to relate the factors to the coordinate system of Fig.\ \ref{fig:dqd}a. We thus write in terms of the spherical coordinates of ${\bf q}$
\begin{align}
A_{{\bf q},l} & = 9 \cos^2(\theta)\sin^4(\theta)\sin^2(2\phi+2\chi), \label{eq:al} \\
A_{{\bf q},t1} & = \tfrac{1}{4}[1+3\cos(2\theta)]^2\sin^2(\theta)\sin^2(2\phi+2\chi), \label{eq:at1}\\
A_{{\bf q},t2} & = \sin^2(2\theta)\cos^2(2\phi+2\chi),\label{eq:at2}
\end{align}
where $\theta = 0$ corresponds to ${\bf q}$ parallel to our $y$-axis, and $\phi$ gives the azimuthal angle of ${\bf q}$ in our $xz$-plane. The angle $\chi$ is the angle between the double dot axis and the crystallographic (100) direction: A wave vector ${\bf q}$ with given $\phi$ thus has an azimuthal angle $\phi+\chi$ in the crystal's coordinate system (see Fig.\ \ref{fig:dir}).

We see that we can write
\begin{equation}
|\lambda_{{\bf q},p}|^2 = \frac{\hbar^2\pi^2v_{p}^2}{q{\cal V}} \left( g_\text{pe}^{(p)}A_{{\bf q},p} + g_{\rm def} q^2  \delta_{p,l} \right),\label{eq:lambda}
\end{equation}
with the two dimensionless coupling constants,
\begin{equation}
g_\text{pe}^{(p)} \equiv \frac{(eh_{14})^2}{2\pi^2\hbar\rho v_{p}^3}
\quad\text{and}\quad
g_{\rm def} \equiv \frac{\Xi^2}{2\pi^2\hbar\rho v_{l}^3a^2},
\end{equation}
the latter being dependent on the dot size $a$.

The relaxation rate $\Gamma$ of the excited $(1,1)$ triplet state to the $(0,2)$ ground state will be calculated using a second order Fermi's golden rule,
\begin{align}
\Gamma =  \sum_f \frac{2\pi}{\hbar}\Bigg| \sum_v \frac{\langle f| \hat H' |v\rangle\langle v|\hat H' | i\rangle }{E_i -  E_v}\Bigg|^2\delta (E_f-E_i),\label{eq:fgr2}
\end{align}
where $\hat H' = \hat H_\text{e-ph} + \hat H_{\rm sm}$, with $\hat H_{\rm sm}$ being one (or both) of the spin-mixing Hamiltonians presented in Sec.~\ref{sec:smpert}. The energy difference $\Delta$ between initial state $\ket{T_+}$ and final state $\ket{S_{02}}$ (which is equal to the energy of the emitted phonon) is assumed much larger than the temperature and we therefore take as initial state a direct product of $\ket{T_+}$ and the phonon vacuum $\ket{{\rm vac}}$, and as final state $\ket{S_{02}}\otimes\ket{1_{{\bf q},p}}$, where one phonon with wave vector ${\bf q}$ and polarization $p$ has been created. 

From the explicit wave functions of $\ket{T_+}$ and $\ket{S_{02}}$ we calculate the diagonal matrix elements $\braket{T_+|\hat H_\text{e-ph}|T_+}$ and $\braket{S_{02}|\hat H_\text{e-ph}|S_{02}}$, and find\cite{endnote3}
\begin{equation}
\Gamma = \sum_{{\bf q},p} \frac{2\pi}{\hbar}|T\lambda_{{\bf q},p}|^2 \left| \frac{ 2F_{\bf q} }{\Delta}+ \frac{F_{\bf q} +e^{i{\bf q}\cdot{\bf d}}F^*_{\bf q} }{-\hbar\omega_{{\bf q},p}}\right|^2\!\delta (\hbar\omega_{{\bf q},p}-\Delta),
\label{eq:gamma1}
\end{equation}
where the spin-mixing matrix element $T=\braket{S_{02}|\hat H_{\rm sm}|T_+}$ and we used the Fourier transform of the squared electronic wave function $F_{\bf q} = \int d{\bf r}\, e^{-i{\bf q}\cdot{\bf r}} |\Psi_{R}({\bf r})|^2$. We can evaluate this Fourier transform explicitly and, anticipating that the $\delta$-function enforces $\hbar\omega_{{\bf q},p} = \Delta$, we write
\begin{align}
 \Gamma = \frac{2\pi}{\hbar} \frac{4}{1-s^2}\sum_{{\bf q},p} & \frac{|T|^2 }{\Delta^2}  | \lambda_{{\bf q},p} |^2 e^{-a^2(q_x^2+q_z^2)} \nonumber\\
 & \times \sin \big(\tfrac{1}{2} q_zd \big)^2 \delta (\hbar\omega_{{\bf q},p} - \Delta ).
\end{align}
The exponential function $\exp\{-a^2(q_x^2+q_z^2)\}$ suppresses the contribution from phonons having a wave vector with in-plane components larger than the inverse system size $1/a$. Indeed, the electronic density profile $F_{\bf q}$ is exponentially small for these wave vectors. The sine function $\sin (\tfrac{1}{2} q_zd )^2$ describes the interference between the coupling to an electron in the left and right dot: A phonon wave with given wave vector ${\bf q}$ has a phase difference $\delta\phi = q_zd$ between the two dot positions.\cite{grangernatphys}

We convert the sum over ${\bf q}$ into an integral and then finally find that we can write for the relaxation rate
\begin{align}
\Gamma = \frac{2\pi}{\hbar}|T|^2 P_{1}(\Delta),
\label{eq:gammapfull}
\end{align}
where the function $P_1(\Delta)$ gives the total probability that the energy $\Delta$ is absorbed by a single phonon, either by piezoelectric coupling or by coupling to the deformation potential. Alternatively, one could call this function the effective phonon density of states for the $(1,1)$ triplet to $(0,2)$ singlet charge transition.

The total single-phonon probability is the sum of the contributions from the different types of phonons,
\begin{equation}
 P_1(\Delta) = P_{1,{\rm def}}(\Delta) + \sum_p P^{(p)}_{1,{\rm pe}}(\Delta). 
\end{equation}
For the piezoelectric phonons we find
\begin{align}
P_{1,{\rm pe}}^{(p)}(\Delta) = \frac{ g_{\rm pe}^{(p)} }{\Delta} \int_0^{\pi/2} \!\!\! d\theta & \, \frac{\sin\theta}{1-s^2} \, f_p \left( \frac{\Delta d\sin\theta}{\hbar v_p} \right) \nonumber\\ 
& \times \exp\left\{-\frac{\Delta^2 a^2}{\hbar^2 v^2_p} \sin^2\theta \right\} ,
\label{eq:p1pe}
\end{align}
where the dimensionless functions $f_p(x)$ are
\begin{align}
f_l(x) = {} & \frac{9}{2} \cos^2(\theta)\sin^4(\theta) g_-(x), \label{eq:fl}\\
f_{t1}(x) = {} & \frac{1}{8}[1+3\cos(2\theta)]^2\sin^2(\theta) g_-(x), \label{eq:ft1} \\
f_{t2}(x) = {} & \frac{1}{2}\sin^2(2\theta) g_+(x), \label{eq:ft2}
\end{align}
in terms of the function
\begin{align}
g_\pm(x) =  {} &  1 - J_0(x)\pm \left(\frac{24}{x^{2}} -1\right) \cos (4\chi) J_0(x) \nonumber\\
& \pm \left(\frac{8}{x} - \frac{48}{x^{3}}\right)  \cos (4\chi) J_1(x),\label{eq:gpm}
\end{align}
with $J_n(x)$ the $n$-th order Bessel function of the first kind. The contribution from the coupling to the deformation potential reads similarly
\begin{align}
P_{1,{\rm def}}(\Delta) = \frac{ g_{\rm def} }{\Delta}\int_0^{\pi/2} \!\!\! d\theta  \,  & \frac{\sin\theta}{1-s^2} \left[ 1-J_0 \left( \frac{\Delta d\sin\theta}{\hbar v_l} \right) \right]\nonumber\\ 
& \times \frac{\Delta^2 a^2}{\hbar^2 v_l^2}\exp\left\{-\frac{\Delta^2 a^2}{\hbar^2 v^2_l} \sin^2\theta \right\}.
\label{eq:p1def}
\end{align}

The remaining integral over the polar angle $\theta$ has to be evaluated numerically. We can however arrive at analytical results in the limits of small and large phonon energies. For small energies, meaning $\hbar v / \Delta \gg a,d$, the single-phonon probabilities $P_{1}$ can be expanded in $\Delta$. Setting $v_{t1} = v_{t2} \equiv v_t$ we then find to leading order
\begin{align}
 P_{1,{\rm pe}}^{(l)}(\Delta) & = \frac{6}{105(1-s^2)} \frac{g_{\rm pe}^{(l)}}{\Delta}\left(\frac{d\Delta}{\hbar v_l}\right)^2,\\
 P_{1,{\rm pe}}^{(t)}(\Delta) & = \frac{8}{105(1-s^2)} \frac{g_{\rm pe}^{(t)}}{\Delta}\left(\frac{d\Delta}{\hbar v_t}\right)^2,\\
 P_{1,{\rm def}}(\Delta) & = \frac{1}{6(1-s^2)} \frac{g_{\rm def}}{\Delta} \left(\frac{a\Delta}{\hbar v_l}\right)^2\left(\frac{d\Delta}{\hbar v_l}\right)^2,
\end{align}
where $P_{1,{\rm pe}}^{(t)} = P_{1,{\rm pe}}^{(t1)}+ P_{1,{\rm pe}}^{(t2)}$. We find for small energies a linear piezoelectric $P_{1,{\rm pe}}(\Delta)$ and a cubic deformation $P_{1,{\rm def}}(\Delta)$, meaning that the phonon bath is superohmic in this setup. The result for the piezoelectric phonons agrees up to a prefactor with previous calculations of the phonon density of states for the $(1,0)$ to $(0,1)$ charge transition where all anisotropy factors were set to one.\cite{PhysRevB.71.205322}

In the opposite limit of large energies, $\hbar v / \Delta \ll a$, we find qualitatively different results compared to Ref.\ \onlinecite{PhysRevB.71.205322}. We see from the exponential factors in (\ref{eq:p1pe}) and (\ref{eq:p1def}) that in this regime only very small angles $\theta$ are relevant. Indeed, in this case only the confinement in the $y$-direction is strong enough to create an electronic density profile with non-vanishing Fourier components of the order $\sim \Delta / \hbar v$, and phonon emission takes place almost exclusively in the $y$-direction. For the piezoelectric coupling the anisotropy factors $A_{{\bf q},p}$ now bring in small factors of $\theta$ which cannot be ignored (see also Ref.\ \onlinecite{khaetskii2}). Since in this limit $A_l \propto \theta^4$ and $A_{t1},A_{t2} \propto \theta^2$, we expect the dominating piezoelectric contribution for large energies to come from transversal phonons.

To evaluate the single-boson probabilities in this limit, we expand $\sin\theta \approx \theta$ and extend the range of integration over $\theta$ from $0$ to $\infty$. Then we find that to leading order
\begin{align}
 P_{1,{\rm pe}}^{(t)}(\Delta) & = \left[ 2 + \frac{d^2}{2a^2}\frac{s^2}{(1-s^2)} \right] \frac{g_{\rm pe}^{(t)}}{\Delta} \left(\frac{\hbar v_t}{a\Delta}\right)^4,\label{eq:p1pele}\\
 P_{1,{\rm def}} (\Delta) & = \frac{g_{\rm def}}{2\Delta}.\label{eq:p1defle}
\end{align}
The contribution from longitudinal piezoelectric phonons is $P_{1,l}\sim {(g_{\rm pe}^{(l)}/\Delta)(\hbar v_l/a\Delta)^6}$, which is much smaller than $P_{1,{\rm pe}}^{(t)}$ and therefore ignored. The large-energy result (\ref{eq:p1pele}) for the piezoelectric phonons is qualitatively different from the results presented in Ref.\ \onlinecite{PhysRevB.71.205322}, which predicted that $P_{1}(\Delta) \sim (g_{\rm pe}/\Delta)(\hbar v/a\Delta)^2$, the difference arising from the inclusion of the anisotropy factors.

\begin{figure}[b]
 \begin{center}
  \includegraphics{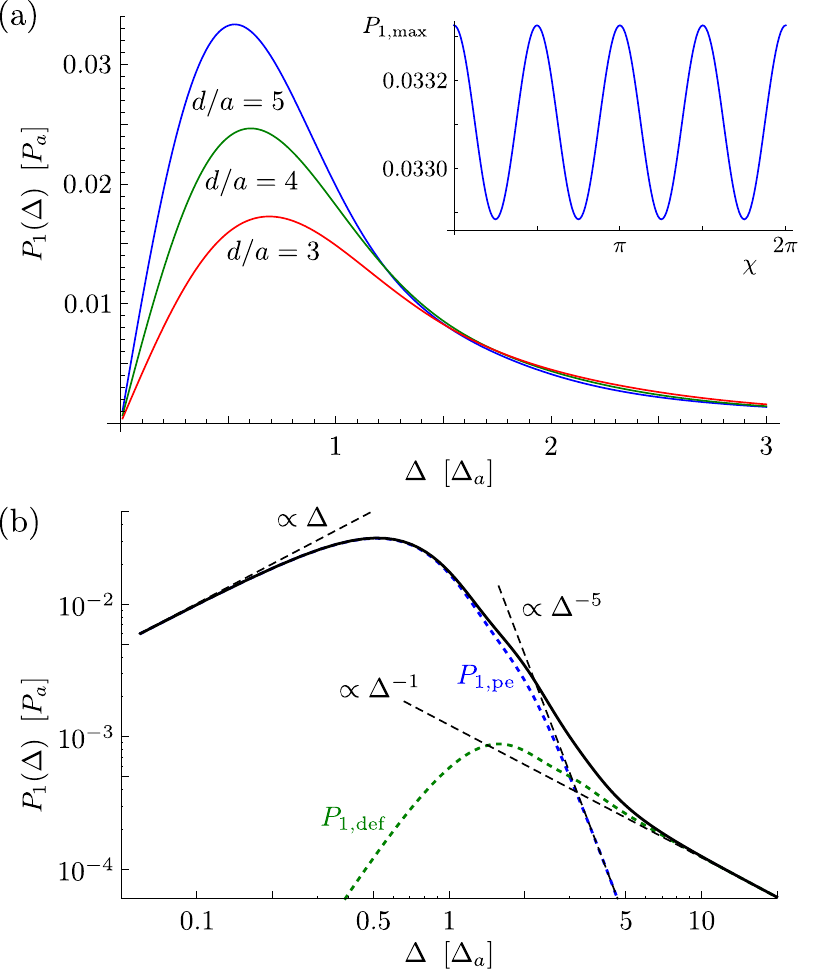}
  \caption{The total single-phonon probability $P_1 = P_{1,{\rm pe}} + P_{1,{\rm def}}$ as a function of $\Delta$. The phonon energy $\Delta$ is plotted in units of $\Delta_a \equiv \hbar v_t / a$ and $P_1$ in units of $P_a \equiv a(eh_{14})^2/\hbar^2v_t^4\rho$. For all plots we have set $v_l/v_t = 1.73$ and $\Xi/aeh_{14} = 0.50$. (a) The total $P_1(\Delta)$ at $\chi=0$ for three different size ratios $d/a$. (inset) The dependence of the maximum $P_1$ for $d/a=5$ on the angle $\chi$. (b) The total $P_1(\Delta)$, as well as the separate contributions from piezoelectric and deformation phonons, for $\chi=0$ and $d/a=5$ on logarithmic scales. The expected power laws are included as guides to the eye.}\label{fig:plots}
 \end{center}
\end{figure}
We see from (\ref{eq:p1pele}) and (\ref{eq:p1defle}) that the contribution from deformation phonons becomes important when $\Delta\sqrt{a} \sim \hbar v_t \sqrt{v_l^3 eh_{14}/v_t^3\Xi}$, which is approximately $1.4$~meV$\cdot{\rm nm}^{1/2}$ using realistic parameters for GaAs.\cite{PhysRevB.48.11144} For a dot size of $a=20$~nm (an orbital level spacing of $\hbar\omega \approx 1.4$~meV) we find that the relevant energy scale is $\Delta \sim 300~\mu$eV, which indeed lies inside our regime of interest.

In Fig.\ \ref{fig:plots} we plot the single-phonon probability $P_1(\Delta)$ for typical double-dot parameters. In all plots $\Delta$ is renormalized to units of $\Delta_a \equiv \hbar v_t / a$ and $P_1$ is plotted in units of $P_a \equiv a(eh_{14})^2/\hbar^2v_t^4\rho$. For $v_t = 3.0\times 10^{3}$~m/s and $a = 20$~nm we find $\Delta_a = 99~\mu$eV. The parameter $v_l/v_t$ was set to $1.73$ and the ratio $\Xi/aeh_{14}$ to $0.50$. In Fig.\ \ref{fig:plots}a we show the total $P_1$ at $\chi = 0$ for three different size ratios $d/a$. The maximum of $P_1$ always occurs on the scale $\Delta \sim \Delta_a$, where the wave length of the emitted phonon is comparable to the system's in-plane dimensions. At low energies $P_1$ is approximately linear and at high energies it is suppressed, ultimately being dominated by the deformation contribution making $P_1 \propto \Delta^{-1}$. In Fig.\ \ref{fig:plots}b we plot the total $P_1$ as well as the two separate contributions on logarithmic scales (for $\chi = 0$ and $d/a=5$), and we added guides to the eye corresponding to the power laws expected in the different limits. The blue dotted line shows the piezoelectric contribution. Up to a few $\Delta_a$ this contribution indeed dominates, being linear at very small energies. For $\Delta \gtrsim \Delta_a$ it becomes suppressed as $\propto \Delta^{-5}$ and at higher energies the dominating contribution comes from the coupling to the deformation potential, the green dotted line (see also Eqs.\ \ref{eq:p1pele} and \ref{eq:p1defle}). The inset of Fig.\ \ref{fig:plots}a shows the dependence of the maximum of $P_1$ on the angle $\chi$ (for $d/a=5$). We see that the density of states depends on the relative orientation of the double dot axis and the (100) crystal direction, its variation being however only $\sim 1$~\%.

\section{Spin-flip charge relaxation rate}\label{sec:rel}

Now we can use Eqs.\ (\ref{eq:gammapfull})--(\ref{eq:gpm}) to evaluate explicit relaxation rates for the $T_+ \to S_{02}$ transition. For a specific experimental setup one can estimate the relative magnitude of the matrix elements given in Eqs.\ (\ref{eq:mso}) and (\ref{eq:mgr}), and decide which process dominates. Here, we will focus on the case of a large external magnetic field, such as was the case in the experiments of Ref.\ \onlinecite{larsnatcomm}. We assume that $B_{\rm ext}$ is large enough so that $|T_{\rm so}| \gg |T_{\rm gr}|$. In that case the relaxation from $\ket{T_+}$ to $\ket{S_{02}}$ is mainly caused by spin-orbit interaction.

We thus use $T_{\rm so}$ in (\ref{eq:gammapfull}) and take typical material parameters for GaAs. For the dot size we take again $a = 20$~nm and the interdot distance is set five times as large, $d = 100$~nm. In the experiment of Ref.\ \onlinecite{larsnatcomm} the interdot axis was fabricated along the crystal (110) direction, so we set $\chi = \pi/4$. For this angle the two spin-orbit terms add constructively, and we choose $\alpha = \beta = 100$~m/s, such that $l_{\rm so} \equiv \hbar/m(\alpha+\beta) \approx 8.6~\mu$m.
\begin{figure}[t]
 \begin{center}
  \includegraphics{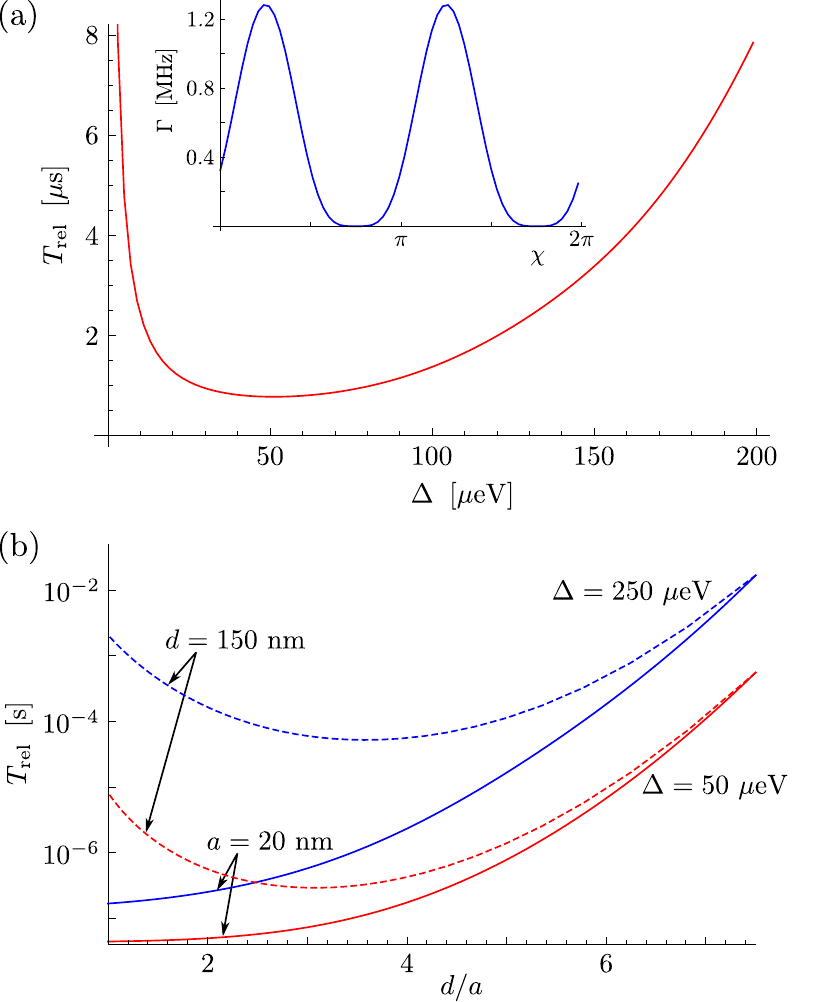}
  \caption{Relaxation time $T_{\rm rel}$ from $\ket{T_+}$ to $\ket{S_{02}}$ when relaxation is mediated by spin-orbit interaction. To make this plot we used $v_l = 5.2\times 10^3$~m/s, $v_t = 3.0 \times 10^3$~m/s, $h_{14} = 1.38 \times 10^9$~V/m, $\rho = 5.3 \times 10^3$~kg/m$^3$, $\chi = \pi/4$, and $\alpha = \beta = 100$~m/s. (a) $T_{\rm rel}$ for $a = 20$~nm and $d = 100$~nm. (inset) The relaxation rate $\Gamma$ for the same parameters calculated at $\Delta = 50~\mu$eV as a function of $\chi$. (b) $T_{\rm rel}$ as a function of $d/a$ at two different energies for fixed $a = 20$~nm (solid lines) and $d = 150$~nm (dashed lines).}\label{fig:plot2}
 \end{center}
\end{figure}
In Fig.\ \ref{fig:plot2}a we plot the resulting relaxation time $T_{\rm rel} = \Gamma^{-1}$, which is found to be typically $T_{\rm rel} \sim 1~\mu$s, the order of magnitude of which agrees with experimental observations.\cite{florist1} The inset to Fig.\ \ref{fig:plot2}a shows the dependence of the relaxation rate $\Gamma$ on the angle $\chi$ at $\Delta = 50~\mu$eV, close to the minimum relaxation time. We see that the rate indeed vanishes for the angles $\chi = 3\pi/4,7\pi/4$, where the Rashba and Dresselhaus terms add destructively.

In Fig.\ \ref{fig:plot2}b we show how the relaxation time depends on the size ratio $d/a$, for two different energies $\Delta = 50~\mu$eV (red lines) and $\Delta = 250~\mu$eV (blue lines). The dashed lines have a fixed interdot distance $d=150$~nm and the solid lines have a fixed dot radius $a = 20$~nm. All other parameters are the same as in Fig.\ \ref{fig:plot2}a. We see that a large size ratio $d/a$ suppresses the relaxation efficiently: For widely separated dots the overlap of the two single-dot wave functions becomes exponentially small, and this suppresses the matrix element $T_{\rm so}$. In the limit of strongly overlapping wave functions, i.e.\ $d/a$ going towards 1, we see that relaxation is much more efficient for the smaller system with $d=a=20$~nm. Indeed, for $d=a=150$~nm we find $\Delta_a \approx 13~\mu$eV, so in this case both energies are larger than $\Delta_a$ where the electron-phonon coupling matrix elements are suppressed.

For smaller external magnetic fields, or other angles $\chi$, one could be in the situation where the spin-orbit and field-gradient give rise to matrix elements of the same order of magnitude, $|T_{\rm so}| \sim |T_{\rm gr}|$. In this case one has to use in (\ref{eq:gammapfull}) the {\it total} matrix element $T_{\rm tot} = T_{\rm so} + T_{\rm gr}$, which possibly includes interference terms between the two mechanisms,
\begin{equation}
 T_{\rm tot} = -i\left\{t\frac{-i\delta B_a^x+\delta B_a^y}{\sqrt{2}B_{\rm ext}}+\frac{\hbar ds( \alpha + \beta\sin2\chi )}{4a^2\sqrt{1-s^2}}\right\}.
\end{equation}
We see that the spin-orbit mechanism interferes with the $y$-component of the difference field $\delta{\bf B}_a$. By tuning $\delta B_a^y$ or $B_{\rm ext}$ one could thus enhance or counteract the spin-flip tunneling enabled by spin-orbit interaction. One word of caution is however required here: If the field gradients are caused by the effective hyperfine fields, then the final state $\ket{f}$ in (\ref{eq:fgr2}) is {\it different} for a spin-orbit and a hyperfine mediated transition. Indeed, in the course of hyperfine induced spin-flip tunneling the spin of one of the nuclei is raised by $\hbar$, which does not happen during a spin-orbit mediated transition. In that case one has to calculate separately the two contributions to the relaxation rate (\ref{eq:gammapfull}) or, equivalently, use $|T|^2 = |T_{\rm so}|^2 + |T_{\rm gr}|^2$.

\section{Conclusion}

We have studied the $(1,1)$ triplet to $(0,2)$ singlet relaxation rate in a lateral gate-defined double quantum dot tuned to the Pauli spin blockade regime. We first derived an effective phonon density of states $P_1(\Delta)$ for this charge transition, and found that at small energies $P_1$ is linear in energy, $\propto \Delta$, and dominated by the piezoelectric electron-phonon coupling, whereas at large energies the $P_1$ is dominated by the coupling to the deformation potential and is $\propto \Delta^{-1}$. Then, we investigated two different spin-mixing mechanisms coupling the spin triplet and singlet states: a magnetic field gradient over the double dot (relevant at low external magnetic field) and spin-orbit interaction (relevant at high field). We showed how the spin-orbit-mediated coupling depends on the device geometry as well as on the in-plane direction of the applied magnetic field. We finally combined all results and took realistic system parameters to evaluate the explicit detuning-dependent relaxation rate, which we found to be of the order of $\sim$~MHz.

\acknowledgments

The author would like to thank F.\ R.\ Braakman, L.\ R.\ Schreiber, L.\ M.\ K.\ Vandersypen, P.\ W.\ Brouwer, and K.\ Flensberg for very helpful discussions.
This work was supported by the Alexander von Humboldt Foundation.


\begin{thebibliography}{10}
\bibitem{endnote1}
Any combination of two odd numbers of excess electrons in the dots leads to the
  same physics. For simplicity we will consider here the most basic situation,
  i.e., two excess electrons in total in the double dot. This situation also
  corresponds to most of the experimental work referred to.
\bibitem{endnote2}
In writing so, we assumed that the difference in Coulomb energy between
  $\ket{S}$ and the $(1,1)$ triplet states can be neglected, which is usually
  realistic when $d \gtrsim a$.
\end{thebibliography}
\end{document}